\documentclass[
 reprint,
nofootinbib,
nobibnotes,
 amsmath,amssymb,
 aps,
]{revtex4-2}

\usepackage{graphicx}
\usepackage{dcolumn}
\usepackage{bm}
\usepackage[whole]{bxcjkjatype}
\usepackage{cleveref}
\usepackage {empheq}
\usepackage{enumerate}
\usepackage{comment}

\crefname{equation}{Eq.}{Eq.}
\crefname{figure}{Fig.}{Fig.}
\crefname{table}{Table}{Table}

\usepackage{tabularx}
\usepackage{array}
\newcolumntype{M}[1]{>{\centering\arraybackslash}m{#1}}
\usepackage[]{xcolor}
\usepackage{colortbl}

\newcommand{\llangle}{\left\langle\!\left\langle}
\newcommand{\rrangle}{\right\rangle\!\right\rangle}

\begin{document}

\preprint{APS/123-QED}

\title{Analysis of discrete modern Hopfield networks in open quantum system}

\author{Takeshi Kimura${}^1$}
\author{Kohtaro Kato${}^1$}
\affiliation{
 ${}^1$Department of Mathematical Informatics,\\ Graduate School of Informatics,\\ Nagoya University, Nagoya 464-0814, Japan
}

\begin{abstract}
The modern Hopfield network, proposed by Krotov and Hopfield, is a mathematical generalization of the Hopfield network, which is a basic model of associative memory that employs higher-order interactions. 
This study introduces an open quantum model for discrete modern Hopfield networks that generalizes the open quantum Hopfield network. 
Our model integrates dissipative quantum spin systems, governed by quantum master equations, with classical hopping terms and additional quantum effects through a transverse field. 
We analytically examined the behavior of the stable fixed points and numerically determined the phase diagram.
The results demonstrated qualitatively distinct behaviors from the open quantum Hopfield network, showing that the ferromagnetic and limit cycle phases have additional stable fixed points.
\end{abstract}

\maketitle

\section{Introduction}
The Hopfield network, introduced by J. J. Hopfield in 1982, is a fundamental mathematical model for associative memory. 
It has been extensively studied in both computer science and statistical physics \cite{hopfield1982neural, PhysRevA.32.1007, PhysRevLett.55.1530, ackley1985learning, gardner1988space, huang2021statistical}. 
In physical terms, a Hopfield network is a spin-glass system with an Ising-like energy function defined on a completely connected graph. 
The network stores input data through coupling constants using the Hebb rule, ensuring that the inputs correspond to the minimum energy spin configurations. 
The stored memories are then retrieved by minimizing the energy function. 

The modern Hopfield network, devised by Krotov and Hopfield, generalizes the energy function of the conventional Hopfield network \cite{krotov2016dense}, significantly increasing its storage capacity~\cite{krotov2016dense, demircigil2017model, ramsauer2020hopfield, lucibello2024exponential}. 
Due to its generality, the modern Hopfield network contains other prominent learning models, such as the transformer~\cite{ramsauer2020hopfield, krotov2021large} used in natural languages, and diffusion models~\cite{hoover2023memory, ambrogioni2024search} used in computer vision, as special cases. 
When discretized, a modern Hopfield network functions equivalently to a neural network with one hidden layer~\cite{krotov2016dense}.

Quantum machine learning, which integrates quantum mechanical effects into machine learning, is actively studied~\cite{biamonte2017quantum}. 
Representative examples include quantum neural networks~\cite{schuld2014quest, PhysRevA.98.032309, farhi2018classification} and quantum kernels~\cite{schuld2019quantum}, which are expected to efficiently explore exponentially large Hilbert spaces.

A key question is whether quantum computers can enhance the classical Hopfield networks.
Several approaches have been proposed to incorporate quantum mechanical effects into Hopfield networks~\cite{schuld2014quest, rebentrost2018quantum, meinhardt2020quantum, rotondo2018open, fiorelli2019quantum, fiorelli2022phase, fiorelli2023mean, bodeker2023optimal}.
One approach involves embedding a Hopfield network into pure quantum states in isolated quantum systems~\cite{schuld2014quest, rebentrost2018quantum, meinhardt2020quantum}, utilizing quantum parallelism to increase the storage capacity and reduce computational complexity.
However, deterministic time evolution often leads to local minima in the energy function.

Another approach involves embedding the network into a dissipative quantum spin system, governed by the quantum master equation~\cite{rotondo2018open, fiorelli2019quantum, fiorelli2022phase, fiorelli2023mean, bodeker2023optimal}. 
Due to stochastic time evolution, these models can avoid stacking at the local minima of the energy function, similar to annealing algorithms. 
In the time evolution of this type of quantum Hopfield network, a transverse field serves as a quantum effect, including non-classical behavior of solutions~\cite{rotondo2018open} and altering storage capacity\cite{bodeker2023optimal}. 
Generalizing this quantum framework to modern Hopfield networks could enhance efficacy and applicability; however, existing studies are limited to extensions of conventional Hopfield networks.

This study generalizes the open quantum Hopfield model to discrete modern Hopfield networks. 
We examine the behavior of stationary solutions of time evolution under random, unbiased memory assumptions. 
We then analyze the phase diagram and storage capacity as in previous models~\cite{rotondo2018open, bodeker2023optimal}. 

In our model, the energy function takes an integer power function, aligning with the Hopfield network when it is quadratic.
Notably, our model exhibits qualitatively distinct behavior when the exponent exceeds two and when the exponent is two (i.e., when the model is an open quantum Hopfield network~\cite{rotondo2018open}).  
When the exponent exceeds two, there is always a stable steady state, and the phase diagram has a region described by a limit cycle (LC) where the origin is always stable. 

The remainder of this paper is organized as follows. 
Section~\ref{sec:model} introduces our proposed model.
Section~\ref{sec:results} presents the findings on the phase diagram, solution behavior, and storage capacity under varying quantum and temperature parameters. 
Section~\ref{sec:conclusion} summarizes our results and discusses future issues.

\section{Open Quantum Discrete Modern Hopfield network} \label{sec:model}

\subsection{Discrete Hopfield Network}
A discrete Hopfield network is defined on a graph consisting of $N$ nodes and edges connecting each node pair.
This model is a recurrent (the output of the node at time $t$ becomes the input at time $t+1$) and asynchronous (update nodes one at a time, and their updates do not affect the other nodes) model, with a fully connected network.
Each node takes the value of $\pm1$, which we interpret as classical 1/2-spins.

Originally developed as a model for associative memory, the Hopfield network retrieves stored spin configurations (memory patterns) using the time evolution of an input configuration.
The time evolution of $i$-th spin is given by:
\begin{align}
    \sigma_i (t+1) &= {\mathrm{sgn}} \Bigg(\sum_{j \neq i} J_{ij} \sigma_j(t) \Bigg), \label{eq:Classical Hopf}\\
    J_{ij} &= \frac{1}{N} \sum_{\mu=1}^p \xi_i^\mu \xi_j^\mu\,,
\end{align}
where $\mathrm{sgn}$ denotes the sign function, $\sigma_i(t) = \pm1 \:(i \in \{ 1, \ldots, N \})$ denotes $N$ binary spin configurations at time $t$, $J_{ij}$ denotes the interaction strength of the edge connecting $i$-th and $j$-th nodes, and $\xi_i^\mu = \pm 1 \:(\mu = \{ 1, \ldots, p \})$ represents $\mu$-th memory pattern of $i$-th spin.

The energy function of Hopfield model is defined as 
\begin{align}
    E &:= - \frac{1}{2} \sum_{i, j} J_{i j} \sigma_i \sigma_j.
\end{align}
The time evolution in Eq.~\eqref{eq:Classical Hopf} decreases the system's energy.
This energy minimization is expected to lead the network to retrieve a memory pattern as the energy eventually minimizes.

\subsection{Discrete modern Hopfield network}
A modern Hopfield network is a mathematical generalization that extends the energy function of the conventional Hopfield network using the function $F$ such that~\cite{krotov2016dense}
\begin{align}
    E = - \frac{1}{2N^{x-1}} \sum_{\mu=1}^{p} F \Bigg( \sum_{j=1}^{N} \xi_j^\mu \sigma_j(t) \Bigg), \quad F(a) = a^x, \label{eq:gen_energy}
\end{align}
where $x$ is typically an integer. 
Similarly, the  time evolution is modified as follows~\cite{krotov2016dense, demircigil2017model}:
\begin{align}
    \sigma_i (t+1) &:= {\mathrm{sgn}} \left( \Delta E_i(t) \right), \\
    \Delta E_i(t) &:= E\left(\sigma_i(t) = 1\right) - E\left(\sigma_i(t) = -1\right) \\
    &= \frac{1}{N^{x-1}} \sum_\mu \xi_i^\mu \Bigg( \sum_{j \neq i}^N \xi_j^\mu \sigma_j(t) \Bigg)^{x-1}\,,
\end{align}
where $E\left(\sigma_i(t) = \pm1\right)$ denotes the energy function with a fixed value of $\sigma_i(t)$. 
The energy function can also be expressed as:
\begin{equation}
    \Delta E_i(t)= \sum_{j_1 \ldots j_{x-1}} W_{i, j_1 \ldots j_{x-1}} \sigma_{j_1}(t) \cdots \sigma_{j_{x-1}}(t),
\end{equation}
using the multivariate interaction coefficients defined as
\begin{align}
    W_{i, j_1 \ldots j_{x-1}} := \frac{1}{N^{x-1}} \sum_{\mu=1}^p \xi_i^\mu \xi_{j_1}^\mu \cdots \xi_{j_{x-1}}^\mu.
\end{align} 
Therefore, although the conventional Hopfield network is a bipartite-interacting model on a complete graph, modern Hopfield networks are multipartite interacting models, complicating analysis.

\subsection{Open quantum generalization of discrete modern Hopfield network}
In this study, we introduce a generalized discrete modern Hopfield network for open quantum systems using the framework proposed in Ref.~\cite{rotondo2018open}.

Let us consider $N$ quantum 1/2-spins (qubits) aligned at the vertices of a completely connected graph of the same size.  
The system evolves under dissipative quantum dynamics, which are described by the quantum master equation:
\begin{align}
    \dot{\rho} = -i [H, \rho] + \sum_{k=1}^N\sum_{\tau=\pm} \left(L_{k\tau} \rho L_{k\tau}^\dagger - \frac{1}{2} \{ L_{k\tau}^\dagger L_{k\tau}, \rho \}\right), \label{eq:qmaster}
\end{align}
where $L_{k\pm}$ are Lindblad operators defined as
\begin{align}\label{def:L}
    L_{k\pm} := \frac{\exp(\pm \beta / 2 \Delta E_k)}{\sqrt{2 \cosh (\beta \Delta E_k)}} \sigma_k^\pm\,,
\end{align}
where $\sigma_k^\pm = (\sigma_k^X \pm i \sigma_k^Y) / 2$ is defined by the Pauli matrices $\sigma^X, \sigma^Y, \sigma^Z$ at site $k$. 
$\beta = 1/T$ is the inverse temperature and 
\begin{equation}
\Delta E_k:=\sum_{j_1 \ldots j_{x-1}} W_{k, j_1 \ldots j_{x-1}} \sigma_{j_1}^Z \cdots \sigma_{j_{x-1}}^Z
\end{equation}
 is the change of ``energy" under flipping the $k$-th spin. 
 Selecting $x=2$ in~\cref{eq:gen_energy} recovers the quantum Hopfield network introduced in Ref.~\cite{rotondo2018open}. 
 The quantum effects of this model are governed by the transverse magnetic field Hamiltonian:
\begin{align}\label{def:H}
    H = \Omega \sum_i \sigma_i^X.
\end{align}

In contrast to the classical models described previously, these quantum models operate at finite temperatures, resulting in stochastic time evolution. 
This stochasticity prevents stacking at local minima, enabling the retrieval of memory through an appropriately scheduled cooling process, similar to annealing algorithms~\cite{doi:10.1126/science.220.4598.671, PhysRevE.58.5355}. 

\subsection{Time evolution of averaged spins}
To further analyze the model, we adopt a mean-field theoretical approach.  
The overlap operator for the $\mu$-th memory pattern and $a$-coordinate components of the spins is defined as follows:
\begin{align}
    M_a^\mu = \frac{1}{N} \sum_{i=1}^N \xi_{i}^\mu \sigma_i^a, \quad a = X, Y, Z.
\end{align}
In the Heisenberg representation, each spin operator evolves over time according to \cref{eq:qmaster}:
\begin{align}
    \dot{\sigma}_i^Z &= -i [H, \sigma_i^Z] + \sum_{k, \tau} \left(L_{k\tau} \sigma_i^Z L_{k\tau}^\dagger - \frac{1}{2} \{ L_{k\tau}^\dagger L_{k\tau}, \sigma_i^Z \}\right), \\
    \dot{\sigma}_i^{\pm} &= -i [H, \sigma_i^\pm] + \sum_{k, \tau} \left(L_{k\tau} \sigma_i^\pm L_{k\tau}^\dagger - \frac{1}{2} \{ L_{k\tau}^\dagger L_{k\tau}, \sigma_i^\pm \}\right). 
\end{align}

By inserting Eqs.~\eqref{def:L} and \eqref{def:H} into this time evolution and ignoring the terms of order $\mathcal{O}(N^{-(x-1)})$, we obtain that~\cite{rotondo2018open}
\begin{align}
    \dot{\sigma}_i^Z &= - \sigma_i^Z + 2 \Omega \sigma_i^Y + \tanh (\beta \Delta E_i) \,,\label{eq:dotsigmaZ}\\
    \dot{\sigma}_i^Y &= - 2 \Omega \sigma_i^Z - \frac{1}{2} \sigma_i^Y.\label{eq:dotsigmaY}
\end{align}
Moreover, the energy difference operator $\Delta E_i$ can be expressed as
\begin{equation}
    \Delta E_i=\sum_{\mu=1}^p\xi^\mu_i(M_Z^\mu)^{x-1}\,.
\end{equation}
By multiplying both sides of Eqs.~\eqref{eq:dotsigmaZ} and \eqref{eq:dotsigmaY} by $\frac{1}{N} \sum_i \xi_i^\mu$, we obtain 
\begin{align}
    \dot{M}_Z^\mu &= - M_Z^\mu + 2 \Omega M_Y^\mu \nonumber\\
    &\qquad+ \frac{1}{N} \sum_{i=1}^N \xi_i^\mu\tanh\left(\beta\sum_{\mu=1}^p\xi^\mu_i(M_Z^\mu)^{x-1}\right), \label{eq:eqmotion0}\\
    \dot{M}_Y^\mu &= -2 \Omega M_Z^\mu - \frac{1}{2} M_Y^\mu. \label{eq:eqmotion}
\end{align}

We assume an unbiased independent and identically distributed random memory pattern ($\xi^\mu_i=\pm1$ with a probability $1/2$). 
Consider the average and the evolution of $\langle M_a^\mu\rangle$, where $\langle \cdot \rangle$ denotes the expected value for randomly stored memory patterns. 
Because the graph is the complete graph, the mean field approximation $\langle M_a^\mu M_b^\nu \rangle \approx \langle M_a^\mu \rangle \langle M_b^\nu \rangle$ becomes exact with the limit of $N\to\infty$~\cite{PhysRevA.32.1007}. 
Similar to Ref.~\cite{rotondo2018open}, we assumed the following self-averaging hypothesis:
\begin{align}
    &\left\langle\frac{1}{N} \sum_{i=1}^N \xi_i^\mu\tanh\left(\beta \sum_\mu \xi^\mu_{i} \left(\langle M_Z^\mu\rangle\right)^{x-1}\right) \right\rangle\nonumber\\
    &\qquad=\llangle \xi^\mu \tanh \left(\beta \sum_\mu \xi^\mu \left(M_Z^\mu\right)^{x-1}\right) \rrangle\,,
\end{align}
where
\begin{align}
    \llangle \cdot \rrangle := \int \mathrm{d} \xi^\mu (\cdot) \mathrm{P}(\xi^\mu),
\end{align}
and $\mathrm{P}(\xi^\mu)$ denotes the probability distribution of randomly stored configuration $\xi^\mu$. 
This assumption is typically valid for disordered systems with the large $N$ limit~\cite{PhysRevA.32.1007}. 
Finally, the equation of motion for the overlaps is obtained as follows:
\begin{align}
    \dot{M}_Z^\mu &= - M_Z^\mu + 2 \Omega M_Y^\mu + \llangle \xi^\mu \tanh (\beta \textstyle\sum_\mu \xi^\mu (M_Z^\mu)^{x-1}) \rrangle, \label{eq:eqmotion0}\\
    \dot{M}_Y^\mu &= -2 \Omega M_Z^\mu - \frac{1}{2} M_Y^\mu. \label{eq:eqmotion}
\end{align}
For simplicity, we omit $\langle\cdot\rangle$ and denote $M_a^\mu\equiv\langle M_a^\mu\rangle$.

\section{Results} \label{sec:results}
This study analyzes the phase diagram of the proposed model. 
The phase diagram of an open quantum Hopfield network is examined in Ref.~\cite{rotondo2018open}, and its storage capacity is analyzed in Ref.~\cite{bodeker2023optimal}. 
We extend these results analytically and numerically, focusing primarily on the case where $x=4$\footnote{We choose $x=4$ to preserve the spin-inversion symmetry of the energy function. Odd values such as $x=3$ break this symmetry and make it difficult to analytically derive the relationship between $\beta$ and $\beta_c$.} and $p=1$. 

\subsection{Phase diagram}
This section presents the phase diagram of our model for $p=1$ (one stored memory). 
Initially, we analyze the number of fixed points in the time evolution of $M_Z$ [Eq.~\eqref{eq:eqmotion0}] and $M_Y$ [Eq.~\eqref{eq:eqmotion}] as follows:
\begin{align}
    \dot{M}_Z &= - M_Z + 2 \Omega M_Y + \tanh (\beta (M_Z)^{x-1}), \label{eq:eqmotion_p1_z}\\
    \dot{M}_Y &= -2 \Omega M_Z - \frac{1}{2} M_Y. \label{eq:eqmotion_p1_y}
\end{align}
Subsequently, we analyze the stability of the fixed points under a small perturbation. 
Finally, we numerically examine the potential presence of LCs in the time evolution.

When $p=1$, the self-consistent equation for the fixed points $\big(\dot{M}_Z = \dot{M}_Y = 0\big)$ is given by:
\begin{align}
    \beta_c M_Z = \tanh(\beta (M_Z)^{x-1}), \quad \beta_c := 1 + 8 \Omega^2. \label{eq:consistency4}
\end{align}
Additionally, we have $M_Y=-4\Omega M_Z$.
Notably, the origin $(M_Z, M_Y) = (0, 0)$ is always a fixed point. 
The boundary at which the number of fixed points changes is determined when the left-hand side and right-hand side of \cref{eq:consistency4} are tangential:
    \begin{empheq} [left = {\empheqlbrace \,}, right = {}]{align}
        & \beta_c M_Z = \tanh(\beta (M_Z)^{x-1}), \nonumber \\
        & \beta_c = \frac{\beta (x-1) (M_Z)^{x-2}}{\cosh^2 (\beta (M_Z)^{x-1})}. \label{eq:mpd}
    \end{empheq}
The parameters $(T,\Omega)$ that satisfy these two equations provide the desired boundary.

\subsubsection{$x=2$}
The phase diagram for the case of $x=2$ was previously studied in Ref.~\cite{rotondo2018open}.
For self-completeness, we briefly reproduce the results in this section. 
From \cref{eq:mpd}, we obtain the relationship between $\beta$ and $\beta_c$ as follows:  
\begin{align}
    \sqrt{1 - \frac{\beta_c}{\beta}} = \tanh \left( \frac{\beta}{\beta_c} \sqrt{1 - \frac{\beta_c}{\beta}} \right), \label{eq:tmp_boundary_x2} 
\end{align}
and the solution is:
\begin{align}
    \beta = \beta_c. \label{eq:boundary_x2}
\end{align}
This defines the boundary where the number of fixed points varies. 
When $\beta<\beta_c$, the origin is the only fixed point; however, when $\beta>\beta_c$, three fixed points exist.

\begin{figure}[h]
  \centering
  \includegraphics[width=\linewidth]{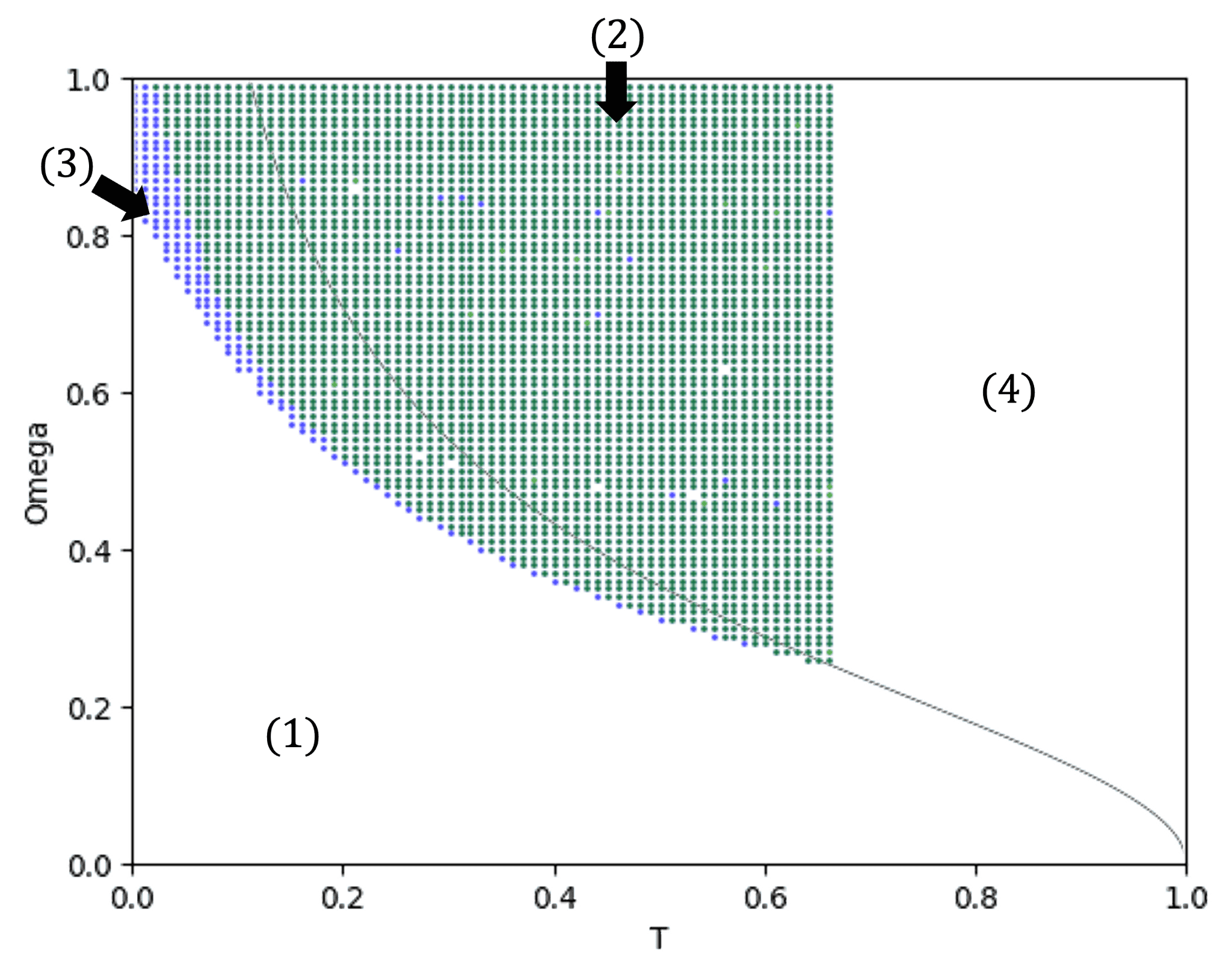}
  \caption{Phase diagram of the quantum modern discrete Hopfield network ($x=2$) in $(T, \Omega)$ plane. 
  The black line indicates the boundary [Eq.~\eqref{eq:boundary_x2}] where the number of solutions changes. 
  The blue+green region signifies the area where limit cycles (LCs) are detected from an initial value away from the origin ($(M_Z, M_Y) = (3, -3)$). 
  The green region is the area where LCs are detected from an initial value around the origin ($(M_Z, M_Y) = (0.05, -0.05)$). 
  Alternatively, the blue region is where the initial values outside the LC converges to it, while the initial value inside the LC and close to the origin converges to the stable fixed points, indicating the ferromagnetic + LC area.
  The numbering from $(1)$ to $(4)$ corresponds to \cref{tb:comparison}.}
  \label{fig:pd2_Hop}
\end{figure}

We also numerically examine the region with the LC, as shown in \cref{fig:pd2_Hop}, where the area above the black line corresponds to a region with a single solution, while the area below has two or more solutions.
However, after examining the stability of each solution, we identify a region below the black line where the ferromagnetic (FM) solution is unstable.
Therefore, we perform a simulation and obtain the green and blue regions, as shown in \cref{fig:pd2_Hop}.
The green region in \cref{fig:pd2_Hop} represents the LC phase, while the blue region is the FM+LC phase. 
Furthermore, the boundary obtained by calculating the stability of the FM solutions in Ref.~\cite{rotondo2018open} does not coincide with the boundary between the blue and green regions in \cref{fig:pd2_Hop}.

\begin{table}[b]
    \caption{Comparison of the phase at $x=2$ and $x=4$.}
    \centering
    \begin{tabularx}{\linewidth}{l >{\centering\arraybackslash}X >{\centering\arraybackslash}X} \hline \hline
         Number & $x=2$ \cite{rotondo2018open} & $x=4$ (this paper)  \\ [6pt] \hline

         $(1)$ & FM & FM \\

         $(2)$ & LC & PM + LC \\

         $(3)$ & FM + LC & FM + LC / PM + LC  \\

         $(4)$ & PM & PM \\ [4pt] \hline \hline
    \end{tabularx}
    \label{tb:comparison}
\end{table}

To assess the stability of the paramagnetic (PM) solution $(M_Z, M_Y) = (0, 0)$, we introduced a small perturbation following Ref.~\cite{rotondo2018open} as:
\begin{equation}
    M_a = 0 + \delta M_a\quad a=Z,Y.
\end{equation}
The time evolution of the perturbed point is given by
\begin{align}
    \delta \dot{M}_Z &= 2 \Omega \delta M_Y - \delta M_Z + \beta \delta M_Z, \label{eq:x2mZ}\\
    \delta \dot{M}_Y &= - 2 \Omega \delta M_Z - \frac{1}{2} \delta M_Y.
\end{align}
The stability matrix of these time evolutions is given as follows:
\begin{equation}
    S =\left(\begin{array}{cc}
        \beta -1 & 2\Omega \\
         -2\Omega & -1/2 
    \end{array}\right).
\end{equation}
Consequently, the stability of the PM solution is given by the following conditions~\cite{rotondo2018open}:
\begin{enumerate}
    \item $\beta > \beta_c$: The eigenvalues are real and different signs. This can be interpreted as the origin being a saddle point.
    \item $\{ \beta < \beta_c \} \cap \{ \beta > 3/2 \} \cap \{ \beta > 4 \Omega + 1/2 \}$: Since the eigenvalues are positive real values, the solution is unstable.
    \item $\{ \beta < \beta_c \} \cap \{ \beta > 3/2 \} \cap \{ \beta < 4 \Omega + 1/2 \}$: Since the eigenvalues have positive real parts and imaginary parts, the solution is unstable and spiraling.
    \item $\{ \beta < \beta_c \} \cap \{ \beta < 3/2 \} \cap \{ |\beta - 1/2 | < 4 \Omega \}$: Since the eigenvalues have negative real parts and imaginary parts, the solution is stable and spiraling.
    \item $\{ \beta < \beta_c \} \cap \{ \beta < 3/2 \} \cap \{ |\beta - 1/2 | > 4 \Omega \}$: Since the eigenvalues are negative real values, the solution is stable.
\end{enumerate}
Conditions 1 and 2 indicate that the PM solution is unstable and are classified as the FM phase.
Condition 3 indicates instability with spiraling behavior and is classified as the LC phase.
Conditions 4 and 5 indicate stability and are classified as the PM phase.

\subsubsection{$x=4$}

When $x=4$, \cref{eq:mpd} implies that
\begin{align}
    &\sqrt{\frac{1 \pm \sqrt{1 - (4 \beta_c^3 / 3 \beta)}}{2}} \nonumber \\
    &\quad = \tanh \left( \frac{\beta}{\beta_c^3} \left( \frac{1 \pm \sqrt{1 - (4 \beta_c^3 / 3 \beta)}}{2} \right)^{3/2} \right), \label{eq:boundary_x4}
\end{align}
which determines the boundary at which the number of fixed points changes. 
When  $\beta>\beta_c$, there are five fixed points instead of three owing to the nontrivial power inside $\tanh(\cdot)$. 

To examine the stability of the PM solution $(M_Z, M_Y) = (0, 0)$, we introduce a small perturbation:
\begin{equation}
    M_a = 0 + \delta M_a\quad a=Z,Y\,.
\end{equation}
The time evolution of the perturbed point is then given by
\begin{align}
    \delta \dot{M}_Z &= 2 \Omega \delta M_Y - \delta M_Z, \\
    \delta \dot{M}_Y &= - 2 \Omega \delta M_Z - \frac{1}{2} \delta M_Y,
\end{align}
where we exclude the terms in $O(\delta M_a^2)$. 
Notably, the term proportional to $\beta$ in~\cref{eq:x2mZ} is negligible. 
The stability matrix of these time evolutions is
\begin{equation}
    S =\left(\begin{array}{cc}
        -1 & 2\Omega \\
         -2\Omega & -1/2 
    \end{array}\right)
\end{equation}
whose eigenvalues are
\begin{align}
    \lambda_{\pm} = \frac{1}{2} \left[ - \frac{3}{2} \pm \sqrt{\frac{1}{4} - 16 \Omega^2} \right].
\end{align}
The real parts of the eigenvalues are always negative, indicating that the origin is always stable.
This phenomenon differs from the quantum Hopfield network (i.e., $x=2$), where the origin can be unstable. 
The time evolution pattern of the perturbed point can be divided into the following cases:
\begin{enumerate}
    \item $| \Omega | \leq 1/8$: The eigenvalues are real numbers and always negative, thereby leading points near the origin to converge directly at the origin.
    \item $| \Omega | > 1/8$: The eigenvalues have negative real values and imaginary values, causing points to spirally converge at the origin.
\end{enumerate}

\begin{figure}[h]
  \centering
  \includegraphics[width=\linewidth]{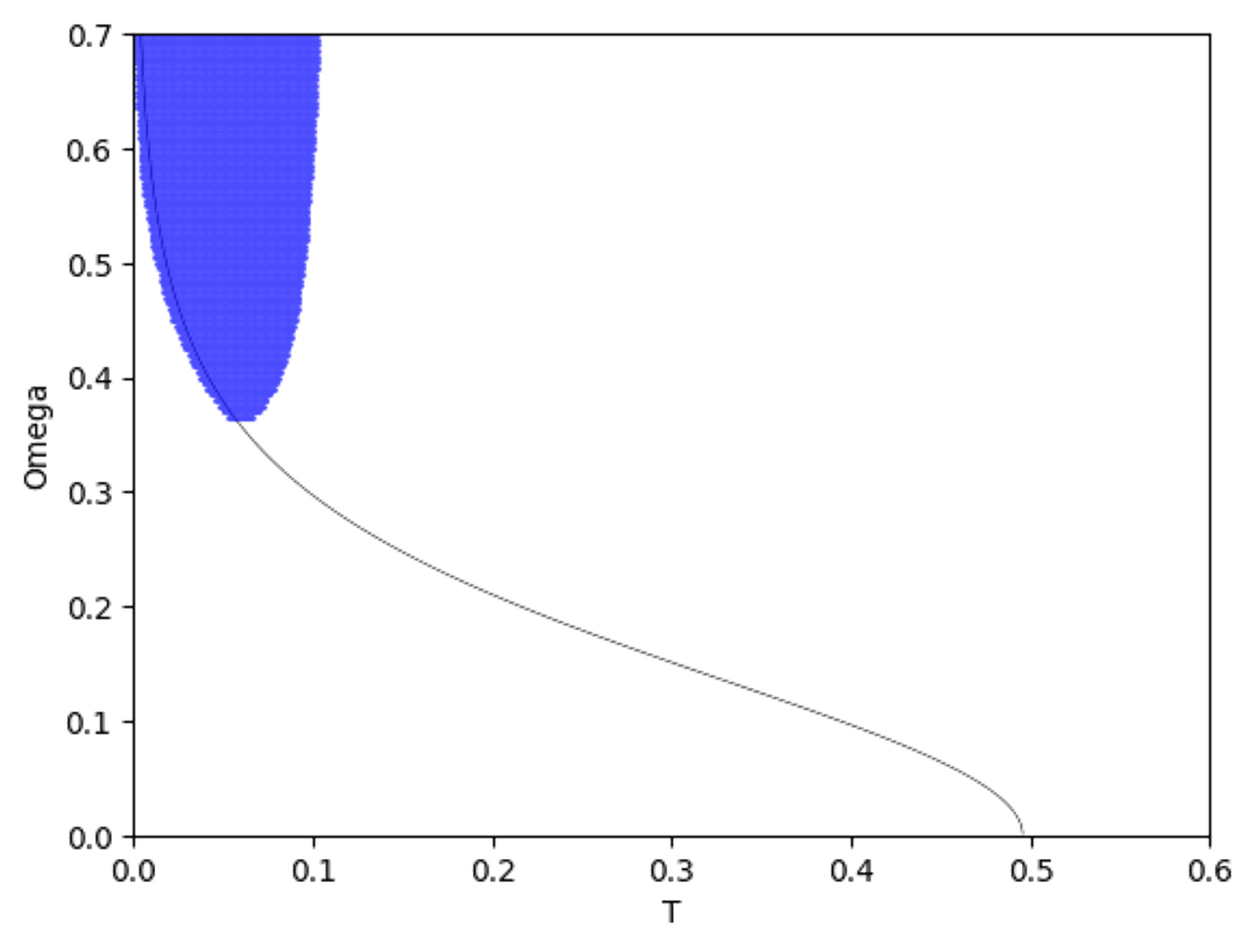}
  \caption{Phase diagram of the quantum modern discrete Hopfield network ($x=4$) in $(T, \Omega)$ plane. The black line indicates the boundary [\cref{eq:boundary_x4}], where the number of solutions changes. 
  The blue region is the area where the simulation suggests the presence of the limit cycle (LC) when the initial value is $(M_Z, M_Y) = (3, -3)$.
  Only one stable attractor in the paramagnetic (PM) phase (above the curve) and the ferromagnetic phase has two or more stable attractors (below the curve). The label LC + PM indicates the region where the LC appears and the origin remains stable.}
  \label{fig:pd1_MoHop}
\end{figure}

\begin{figure}[h]
  \centering
\includegraphics[width=\linewidth]{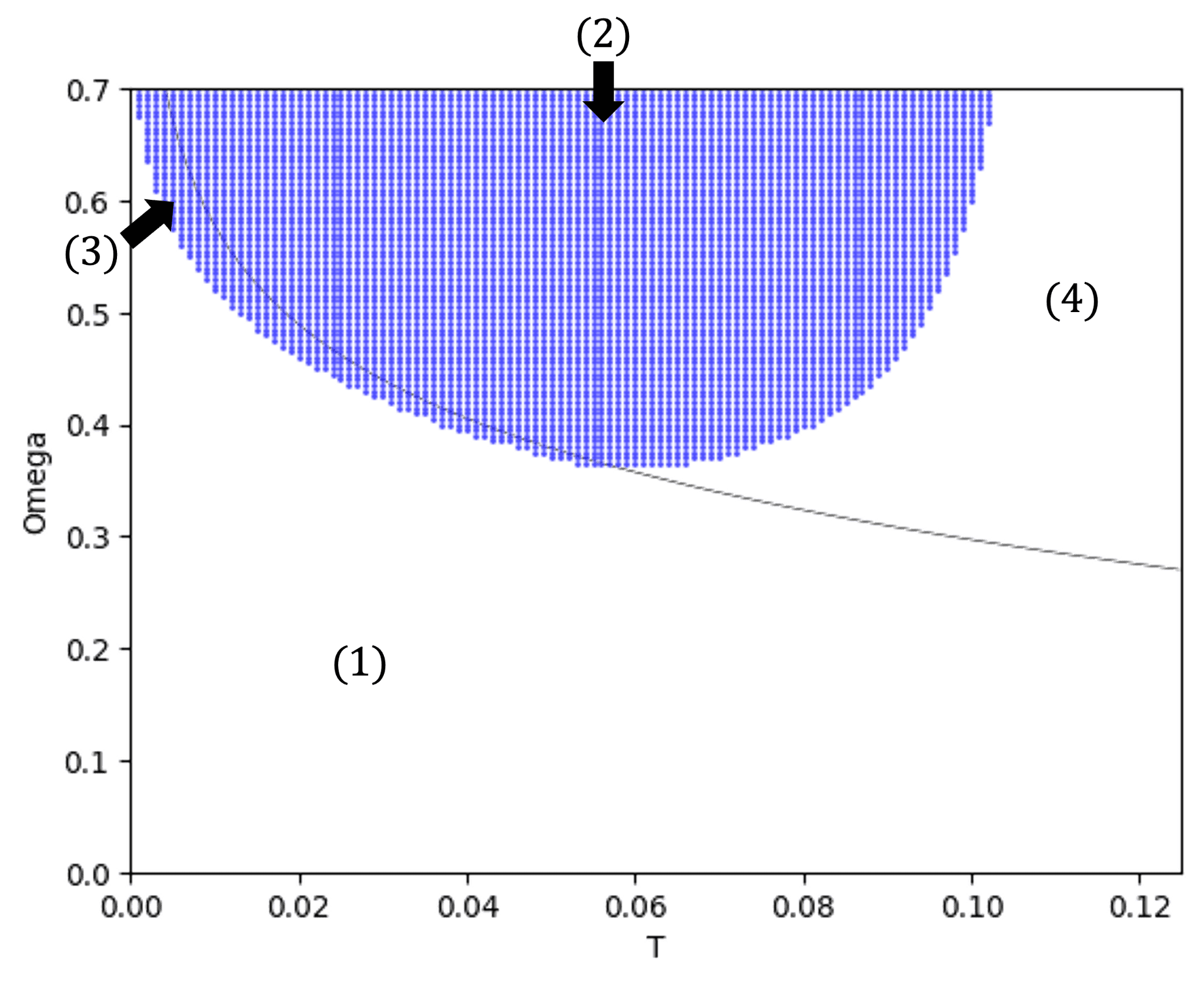}
  \caption{Enlarged view of the region where the limit cycle (LC) appears. $(1)$ to $(4)$ represent the number in \cref{tb:comparison}, $(2)$ refers the blue region above the black line and $(3)$ refers to the narrow blue region below the black line. The region $(3)$ contains two different phases, depending on whether the fixed points with $M_Z>0$ are stable (ferromagnetic+LC) or unstable (paramagnetic+LC).}
  \label{fig:pd2_MoHop}
\end{figure}

Subsequently, we analyze the stability of other fixed points (not the origin).
Let us consider a small perturbation around the solution $\overline{M}_a$:
\begin{align}
    M_a = \overline{M}_a + \delta M_a\,.
\end{align}
We obtain the following equations for the $O(\delta M_a)$ terms:
\begin{align}
    \delta \dot{M}_Z &= 2 \Omega \delta M_Y - \delta M_Z + \frac{3 \beta (\overline{M}_Z)^2}{\cosh^2 (\beta (\overline{M}_Z)^3)} \delta M_Z, \\
    \delta \dot{M}_Y &= - 2 \Omega \delta M_Z - \frac{1}{2} \delta M_Y.
\end{align}
By replacing $\beta$ with the stability of the PM solution of $x=2$, we obtain
\begin{equation}
    \beta \to \beta' = \frac{3 \beta (\overline{M}_Z)^2}{\cosh^2 (\beta (\overline{M}_Z)^3)}.
\end{equation}
This transformation allows us to obtain the stability around the solution, classifying it similarly to the case for $x=2$.
However, this boundary changes slightly from the existing boundary.

In addition to the fixed points, we numerically identify some parameter regions with a nontrivial LC encircling the origin (blue region in Figs.~\ref{fig:pd1_MoHop} and~\ref{fig:pd2_MoHop}). However, unlike in the open quantum Hopfield network, the LC consistently coexists with at least one stable fixed point (origin), and the shape of the region in the phase space differs qualitatively from that of the quantum Hopfield network.

\subsection{Classification of the regions in the phase diagram ($x=4$)}
In this section, we present the various phases of our model in the $(T,\Omega)$ phase space.

\subsubsection{PM phase}
The PM phase corresponds to the region where every initial state reaches the origin, which is the only stable fixed point ((d) and (e) in Fig.~\ref{fig:pd3}).

\subsubsection{PM+LC phase}
The PM+LC phase corresponds to the region where the only stable fixed point is the origin; however, there is an additional LC around it ((a), (b) and (c) in Fig.~\ref{fig:pd4}). 
In a quantum Hopfield network, the LC phase does not coexist with the PM phase; thus, any initial $(M_Z, M_Y)$ converges to the LC. 
However, in our model, the LC always coexists with other stable fixed points.
In the PM+LC phase, the initial point converges to the origin or the LC depending on its proximity to the origin.

\subsubsection{FM phase}
\begin{figure}[h]
  \centering
  \includegraphics[width=\linewidth]{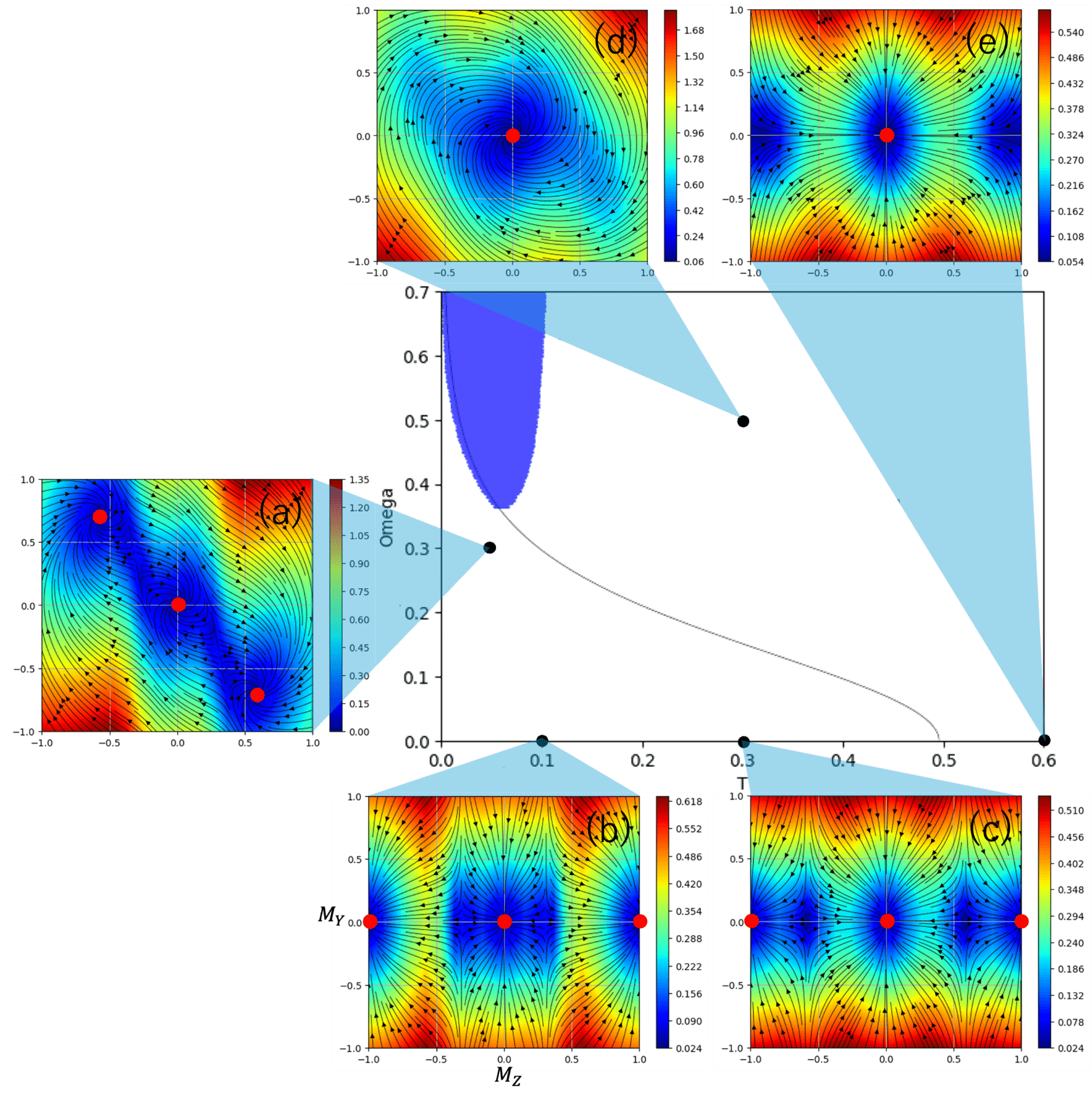}
  \caption{The time evolution of the averaged spin for each parameter $T, \Omega$ in the ferromagnetic (FM) and paramagnetic (PM) phases. The red dots are the stable fixed points. (a), (b) and (c) represent the time evolution in the FM phase, and (d) and (e) represent the time evolution in the PM phase.}
  \label{fig:pd3}
\end{figure}

The region with stable fixed points, such that $M_Z > 0$, but without the LC, is termed the FM phase ((a), (b) and (c) in Fig.~\ref{fig:pd3}).
Unlike the FM phase of a quantum Hopfield network~\cite{rotondo2018open}, which only has stable fixed points with $M_Z>0$, our model exhibits the FM phase with an additional stable fixed point at the origin. 
Consequently, if the initial $M_Z$ is sufficiently small, the magnetization $M_Z$ diminishes, similar to the PM phase.
In addition, the FM region in \cref{fig:pd1_MoHop} is smaller than that in a quantum Hopfield network~\cite{rotondo2018open}.

\subsubsection{FM+LC phase}
\begin{figure}[h]
  \centering
  \includegraphics[width=\linewidth]{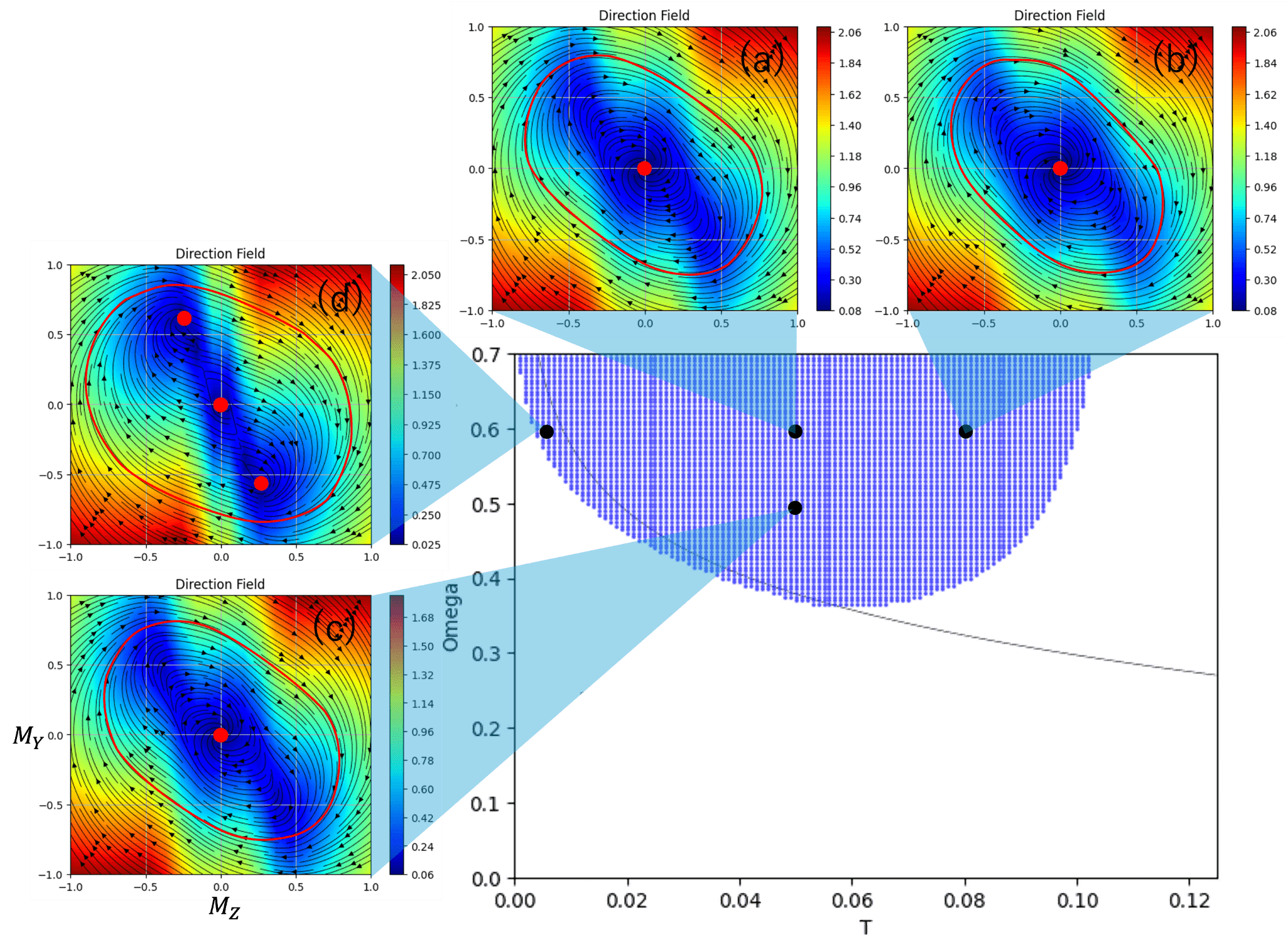}
  \caption{The time evolution of the averaged spin for each parameter $T, \Omega$ in the phase in which the limit cycle (LC) appears. (a), (b) and (c) represent the time evolution of the averaged spin in the paramagnetic+LC phase, and (d) represents the time evolution of the averaged spin in the ferromagnetic+LC phase.}
  \label{fig:pd4}
\end{figure}

The FM+LC phase corresponds to the region where FM and LC coexist ((d) in Fig.~\ref{fig:pd4}). 
In this region, LC encapsulates all stable fixed points. 
When the initial point lies within the LC, it converges to one of the fixed points or the LC, depending on the closeness to these points.
Conversely, any point outside the LC converges to the LC. 

\subsubsection{Behavior of the fixed points for larger $x$}
\begin{figure}[h]
  \centering
  \includegraphics[width=\linewidth]{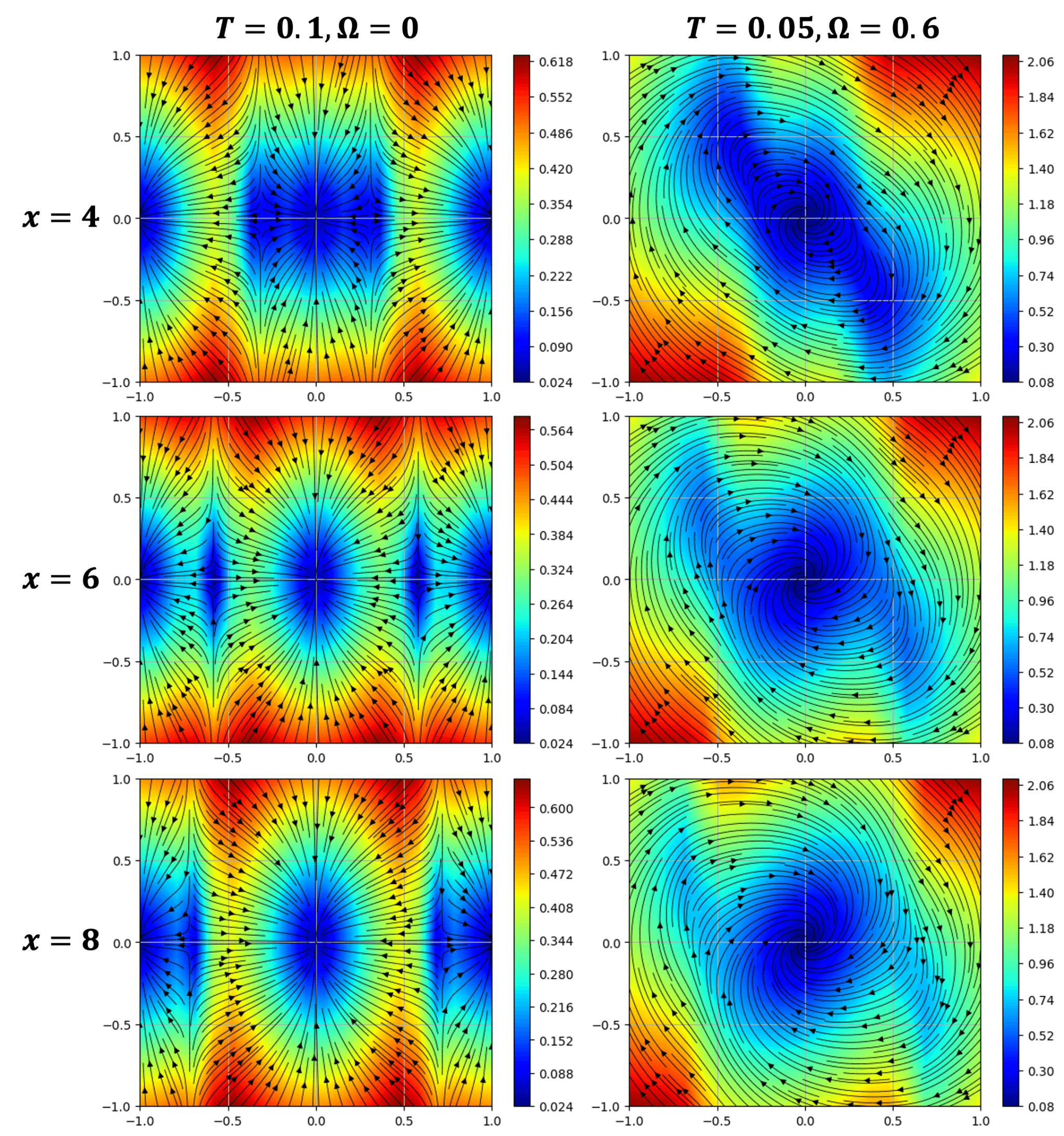}
  \caption{The time evolution of the averaged spin as the value of $x$ increases. As the value of $x$ is increased, the range of $M_Z$ that converges to the origin becomes larger.}
  \label{fig:largex}
\end{figure}

We also analyze the time evolution for $x$ values exceeding $4$ (~\cref{fig:largex}). 
In a discrete modern Hopfield network, memory capacity increases nonlinearly with $N$ as $x$ increases~\cite{krotov2016dense}. 
The time evolution of the averaged spins varies with increasing $x$ in an open quantum discrete Hopfield network.
The point of change is that the range of $M_Z$ that converges to the origin expands as the value of $x$ increases.

The open quantum discrete modern Hopfield network is described by Eqs.~\eqref{eq:eqmotion_p1_z} and \eqref{eq:eqmotion_p1_y}, and the behavior of these functions changes only with $x$ via $\tanh (\beta (M_Z)^{x-1})$.
The function $\tanh (\beta (M_Z)^{x-1})$ has no extreme value at $x=2$, but has an extreme value $M_Z = 0$ when $x>2$, indicating that the origin is always stable.
Consequently, the behavior of the model changes qualitatively when moving from $x=2$ to $x>2$, but remains qualitatively unchanged for different values of $x>2$.

\subsection{Storage capacity}
In the problem of associative memory, storage capacity is the maximum number of patterns that can be retrieved.
When considering a landscape of energy function, each stored pattern is an attractor of that landscape.
Two methods are used to analyze the storage capacity of Hopfield networks.
The first calculates the maximum number of patterns that can be stored without unacceptable errors~\cite{hopfield1982neural, 10.5555/574634}.
The second employs statistical physics analysis~\cite{gardner1988space, 10.5555/574634}.
Notably, the Hopfield network can store $0.14 N$ patterns.

The discrete modern Hopfield network can store $\alpha_x N^{x-1}$ patterns where $\alpha_x$ is a numerical constant that depends on the threshold of error~\cite{krotov2016dense}.
Moreover, models with more generalized energy functions and time evolution can achieve an exponential storage capacity~\cite{demircigil2017model, ramsauer2020hopfield}. 
The storage capacity $\alpha_c$ of a quantum Hopfield network at a given temperature $T$ and quantum drive $\Omega$ is also known~\cite{bodeker2023optimal} (note that $\alpha_c$ is the number of patterns per node, which is sometimes called the storage capacity).
The storage capacity of our model can be calculated using the method described in Ref.~\cite{bodeker2023optimal}.
While the classical and quantum limits coincide, further exploration of quantum capacity remains a future challenge.

\section{Conclusion} \label{sec:conclusion}
In this study, we propose an open quantum discrete modern Hopfield network, a generalization of the open quantum Hopfield network, and determine its time evolution and phase diagram.
Unlike the conventional Hopfield model ($x=2$), the origin ($M_Z=0$) is always stable in our model ($x=4$), and the region with the LC is smaller. 
Each phase is determined by combining the analytical calculation of the number of fixed points with numerical analysis of the LC’s existence.
Although our model qualitatively differs from the open quantum Hopfield model ($x=2$), the behavior of $x>2$ does not change significantly. 

As the classical modern Hopfield networks have broad applications and large capacity, we expect that our quantum variant will offer similar advantages.
However, the strong nonlinearity makes the interactions multivariate, making the analytical calculation of the capacity challenging. 
The memory capacity for the open quantum Hopfield model is calculated through the dilution of the network and cumulant expansion; however, the applicability of these tools to our model remains unclear. 

As future directions, we propose two promising extensions of our model that aim to enhance its realism and applicability. First, one can introduce interactions between each node and its local environment to model the effects of noise. This is particularly relevant for realistic implementations, where local environmental interactions are unavoidable. Studying how such noise influences the system's behavior could offer valuable insights into the model's robustness under practical conditions. Second, a natural extension is to explore time-dependent coupling to the system, particularly in the context of quantum annealing. Since our framework ultimately aims at energy minimization, implementing an annealing schedule, where the temperature is gradually lowered over time, would be a compelling direction. In this context, identifying effective annealing protocols becomes a central issue, especially for practical use cases. 

In addition to these discrete-variable models, another important direction is to generalize the framework to continuous variable systems. Key subclasses of classical modern Hopfield networks, such as transformer and diffusion models, are formulated in this way. Investigating quantum analogs of these continuous architectures remains a challenging and significant goal for future research.

\begin{acknowledgments}
T.K. acknowledges support from MEXT Q-LEAP JPMXS0120319794 and MEXT Q-LEAP JPMXS0120351339; from JST SPRING, Grant No. JPMJSP2125, and would like to thank the “THERS Make New Standards Program for the Next Generation Researchers.''
K.K. acknowledges support from JSPS KAKENHI Grant-in-Aid for Early-Career Scientists (Grant No. JP22K13972), Grant-in-Aid for Transformative Research Areas (B) (Grant No. JP24H00829), and Grant-in-Aid for Scientific Research (C) (Grant No. JP23K17668).
\end{acknowledgments}

\nocite{*}

\bibliography{ref}

\end{document}